\documentclass[doublecol,figures]{epl2}

\usepackage{amssymb,amsmath}
\usepackage[T1]{fontenc}
\def\prb{Phys.\ Rev.\ B}
\def\prl{Phys.\ Rev.\ Lett.}

\def\epl{Europhys.\ Lett.}
\def\apl{Appl.\ Phys.\ Lett.}
\def\epjb{Eur.\ Phys.\ J.\ B}

\title{Magnetic ordering in the striped nickelate La$_{5/3}$Sr$_{1/3}$NiO$_4$:\\
A band structure point of view}

\author{ Udo Schwingenschl{\"o}gl\inst{1} \and 
         Cosima Schuster\inst{1} \and
         Raymond Fr\'esard\inst{2}  }

\shortauthor{U. Schwingenschl{\"o}gl \etal}
\shorttitle{ Band structure of striped nickelates }

\institute{
  \inst{1} TP II, Institut f{\"u}r Physik, Universit\"at Augsburg, D-86135
  Augsburg, Germany \\
  \inst{2} Laboratoire CRISMAT, UMR CNRS--ENSICAEN(ISMRA) 6508,
           6 Bld. du Mar\'echal Juin, F-14050 Caen, France }


\abstract{
We report on a comprehensive study of the electronic and magnetic structure
of the striped nickelate La$_{5/3}$Sr$_{1/3}$NiO$_4$. The investigation is
carried out using band structure calculations based on density functional
theory. A magnetic structure compatible with experiment is obtained from
spin-polarized calculations within the generalized gradient approximation (GGA),
whereas inclusion of a local Coulomb interaction in the LDA+$U$ framework 
results in a different ground state. The influence of the various interaction
parameters is discussed in detail.
}

\begin{document}

\maketitle

Stripe phases are currently the focus of an intense activity as they
are observed in a wealth of systems. Most of them are layered oxides,
including cuprates \cite{Tra96Nd,Abb05,Koh07,Li07},
nickelates \cite{Lee97,Yos00,Lee02,Fre02,Hue06}, and
manganites \cite{Dag03,Kim02,Lar05}. In such doped Mott insulators
the structure of these phases varies from system to system: the domain walls
in which the holes are primarily located can run parallel to the lattice 
axes, or along the diagonal. Furthermore, the distance between them is either
compatible with one doped hole per two domain walls, in the so-called
\textit{half-filled stripes}, or with one doped hole per one domain
wall in the \textit{filled stripes}, as discussed by Zaanen and Ole\'s
\cite{Zaa96}. Those various structures are realized in different systems,
e.\ g., half-filled vertical stripes in the lightly doped cuprate
Ca$_{1.88}$Na$_{0.12}$CuO$_2$Cl$_2$, and diagonal filled stripes in
La$_{2-x}$Sr$_x$NiO$_4$. Stripe formation manifests
itself in a wealth of experiments, in particular by the appearance of
peaks in both unpolarized and polarized neutron scattering experiments.
Indeed charge and spin order in La$_{2-x}$Sr$_x$NiO$_4$  is  
characterized by the wave vectors ${\bf Q}_{c}=2\pi(\epsilon,\epsilon)$ 
and ${\bf Q}_{s}=\pi(1\pm\epsilon,1\pm\epsilon)$, respectively, with
$\epsilon\simeq x$ corresponding  to a constant charge density of one hole/Ni
ion along the diagonal stripe.
\begin{figure}
\begin{center}
\includegraphics[width=0.35\textwidth,clip]{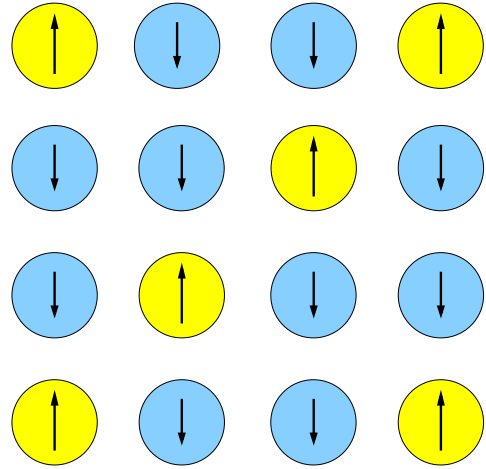}
\end{center}
\caption
{
(Color online) 
Magnetic structure of the Ni atomes in the NiO-planes of
La$_{5/3}$Sr$_{1/3}$NiO$_4$. Bright (yellow)
circles correspond to Ni1/Ni2 sites, dark (blue) circles
to Ni3/Ni4 sites. The representation refers to the experimental
CDFAS phase, see the text for details.
}
\label{fig:fig0}
\end{figure}
Incidentally, the electronic states involved in the stabilization of the stripe
phases form \textit{mid-gap} bands; in the first case these bands are 
partially filled, sometimes even half-filled, while in the second case these
bands are filled (for a recent discussion see Ref.~\cite{Rac07}).  From the point of view of
describing these phases by means of microscopic models it is commonly accepted that
the one band Hubbard model is suitable for the cuprates, whereas the orbital
degeneracy incorporated in a two-band Hubbard Hamiltonian plays a prominent
role when studying nickelates. In both cases it turns out that the
parameters of the model deeply influence the type of ground state
(see, e.\ g., Ref.~\cite{Rac06epl}), and it
would be desirable to obtain an estimate of them from \textit{ab initio}
calculations. Indeed, when comparing a doubly degenerate Hubbard model to a
more  realistic Hamiltonian containing e$_g$ states \cite{Rac06n}, the former
predicts bond centered diagonal half-filled stripes to be the ground state,
whereas diagonal filled stripes are promoted in the latter case. Besides, the relevance
of the Jahn-Teller coupling was emphasized by Zaanen and Littlewood
\cite{Zaa94}. Therefore a band structure calculation embracing all these
aspects is expected to provide insight into the mechanism leading to the stripe
formation. As compared to the cuprates, quantum fluctuations should play a
smaller role in the nickelates where the Ni$^{2+}$ ions carry spin 1.
The approximations involved in the band structure calculations hence are expected to
have less impact. In this work we focus on the doping $x = 1/3$, where the
wave vectors characterizing spin and charge modulations coincide as
$Q_{s,c}=\frac{\pi}{3}(1,1)$. This case corresponds to the most stable stripe
phase with an ordering temperature of $T\sim240$\,K, as revealed by specific heat
\cite{Ram96}, transport \cite{Che94}, and optical conductivity data
\cite{Kat96}. For a more comprehensive review of the experimental situation we
refer to Ref.~\cite{Rac06n}.

\begin{table}
\begin{center}
\begin{tabular}{l||c|c}
&semi-core states&valence states\\\hline
Sr & $4s$, $4p$ & $5s$, $5p$\\
Ni & $3p$ & $3d$, $4s$, $4p$\\
La & $5s$, $5p$ & $4f$, $5d$, $6s$, $6p$\\
O & $2s$ & $2p$, $3s$
\end{tabular}
\end{center}
\caption{\rm Choice of semi-core and valence states.}
\label{tab2}
\end{table}
\begin{figure}
\includegraphics[width=0.48\textwidth,clip]{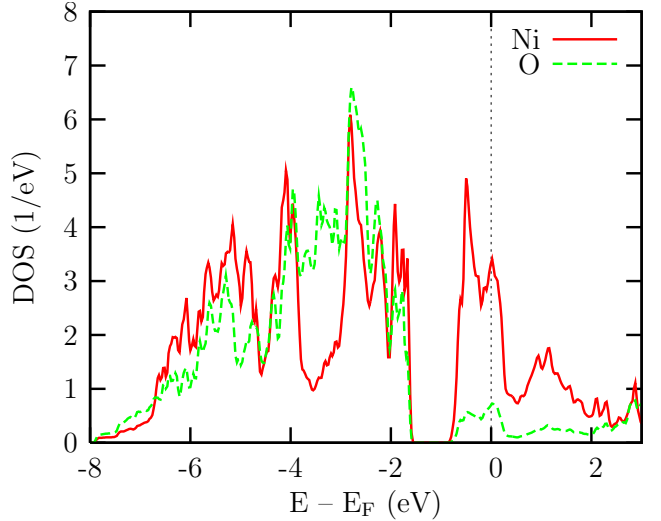}
\caption
{
(Color online) 
Partial Ni $3d$ and O $2p$ DOS (per unit cell)
for spin-degenerate La$_{5/3}$Sr$_{1/3}$NiO$_4$, as obtained by the LDA+$U$ method.
}
\label{fig:fig1}
\end{figure}

The crystal structure of the La$_{2-x}$Sr$_x$NiO$_4$ compounds is based on the
body-centered tetragonal K$_2$NiF$_4$ structure with space group $I4/mmm$
\cite{takeda90}, as for the La$_{2-x}$Sr$_x$CuO$_4$ system \cite{nguyen}.
Therefore, the tetragonal unit cell comprises two La$_{2-x}$Sr$_x$NiO$_4$
formula units. The compounds essentially consist of NiO$_6$ octahedra forming
layers parallel to the $ab$-plane, where neighbouring layers are separated
by La/Sr ions. To be more specific, the magnetic Ni atoms are arranged on a square
planar lattice and interact via the O atoms located midway between them. By symmetry,
there are two crystallographically inequivalent NiO planes, which, however,
are closely related to each other. We have obtained the structural parameters
for our band structure calculations by interpolation between the experimental
values at compositions of $x=0$ and $x=0.6$. For the composition of interest,
$x=1/3$, the lattice constants amount to $a=3.83$\,\AA\ and $c=12.68$\,\AA.
While the La/Sr sites are located at (4e) Wyckoff positions with parameter
$z=0.3617$, the Ni and O(1) sites occupy (2e) and (4c) Wyckoff positions,
respectively. Finally, for the O(2) sites we again have to deal with (4e)
Wyckoff positions, where the parameter is $z=0.174$. In order to account
for the realized spin and charge modulations, we set up a supercell of the
tetragonal unit cell. Comprising altogether six La$_{5/3}$Sr$_{1/3}$NiO$_4$
formula units with 12 La/Sr sites, our supercell is given by the lattice
vectors $(1,1,0)$, $(3,0,0)$, and $(0,0,1)$, with respect to the tetragonal lattice.
In order to satisfy the La:Sr ratio of 5:1, we have to attribute 10 La and 2 Sr atoms
to the 12 La/Sr sites, where the details of the distribution do not affect
our further argumentation. As a consequence, we have four classes of inequivalent
Ni sites, called Ni1, Ni2, Ni3, and Ni4 in the following. Sites Ni1 and Ni3
are located in one of the inequivalent NiO planes, while sites Ni2 and
Ni4, respectively, correspond to them in the other NiO plane. The positions
of sites Ni1 and Ni3 are translated into that of Ni2 and Ni4, respectively,
by the vector $(a/2,a/2,c/2)$. The charge and
spin pattern covered by these structural prerequisites is illustrated in
fig.\ \ref{fig:fig0}. We have diagonal stripes of Ni1 (Ni2) sites, separated by
domains of Ni3 (Ni4) sites, thus reflecting the coincidence of the spin
and charge wave vectors.

The electronic structure calculations presented in the following rely on
density functional theory (DFT). Concerning the DFT implementation, we
apply the Wien2k package, a state-of-the-art full-potential code with
mixed lapw and apw+lo basis set \cite{wien2k}. This code has shown great
capacity in dealing with the interplay of structural relaxation,
magnetic interaction and electronic correlations in very complex materials
\cite{wien2kuscs}. In each of our calculations, the charge density is
represented by almost 23,000 plane waves and the mesh for the Brillouin
zone integration comprises 60 {\bf k}-points in the irreducible wedge.
Our choice of semi-core and valence states is summarized in table
\ref{tab2}. For the supercell setup described above, we address calculations
within the pure generalized gradient approximation (GGA) as well as
LDA+$U$ calculations to account for an additional local electron-electron
interaction. We use the Perdew-Burke-Ernzerhof \cite{perdew} parameterization
of the exchange-correlation functional for the pure GGA, while the LDA+$U$
calculations are based on the SIC scheme described in \cite{anisimov,liechtenstein}.
In each case, we use a band structure calculation
where we artificially enforce spin-degeneracy as reference for the
total energy gain of specific spin patterns. In particular, we investigate
diagonal filled ferromagnetic stripes (DFFS) as well as A- and C-type
diagonal filled antiferromagnetic stripes (ADFAS, CDFAS), which allows us to
compare ferromagnetic with antiferromagnetic coupling both within
the NiO planes (intraplane) and between adjacent planes (interplane).

The electronic states at the Fermi energy are expected to grow out of the
Ni $3d$ and O $2p$ orbitals, due to Ni-O bonding. This is confirmed by
fig.\ \ref{fig:fig1}, which shows the Ni and O partial densities of states
(DOS) resulting from a spin-degenerate LDA+$U$ calculation. Contributions
of La and Sr states in the energy interval of fig.\ \ref{fig:fig1} are very
small and play no role for the further discussion. Here, and
in the following LDA+$U$ calculations, local interactions are assumed for
both the Ni$^{2+}$ and Ni$^{3+}$ sites. For the onsite Coulomb interaction
we choose a value of $U=8$\,eV and for the Hund's coupling a value of
$J_H=0.8$\,eV \cite{anisimov91,anisimov99}. In fig.\ \ref{fig:fig1} we
identify two groups of electronic bands, the first extending from about $-8$\,eV
to $-1.6$\,eV, with respect to the Fermi energy. The second, being separated
from the first group by an energy gap of 0.7\,eV, starts at about $-1.9$\,eV
and extends far beyond the Fermi level. Whereas O $2p$ contributions concentrate
in the low energy group of bands, the DOS at higher energy is dominated by the
Ni $3d$ states. In particular, a pronounced DOS structure is evident at the
Fermi energy, which points towards an instability against magnetic ordering.

\begin{figure}
\includegraphics[width=0.48\textwidth,clip]{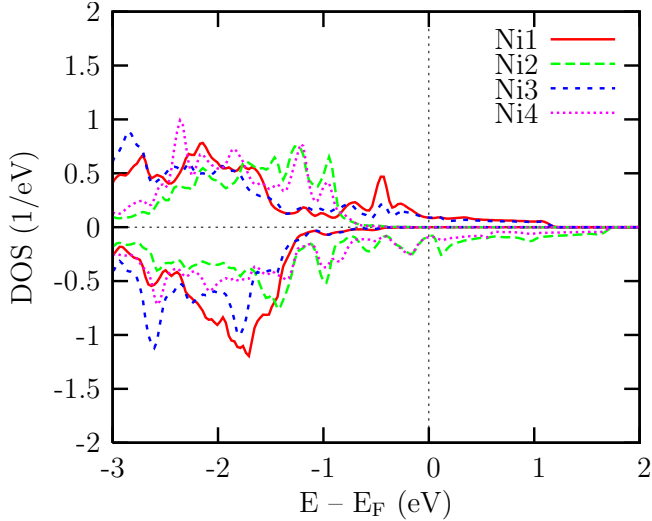}
\caption
{
(Color online) 
Partial spin-majority and spin-minority Ni $3d$ DOS (per Ni atom)
for spin-polarized La$_{5/3}$Sr$_{1/3}$NiO$_4$ in the
ADFAS phase, as obtained by the LDA+$U$ method.
}
\label{fig:fig2}
\end{figure}

\begin{figure}
\includegraphics[width=0.48\textwidth,clip]{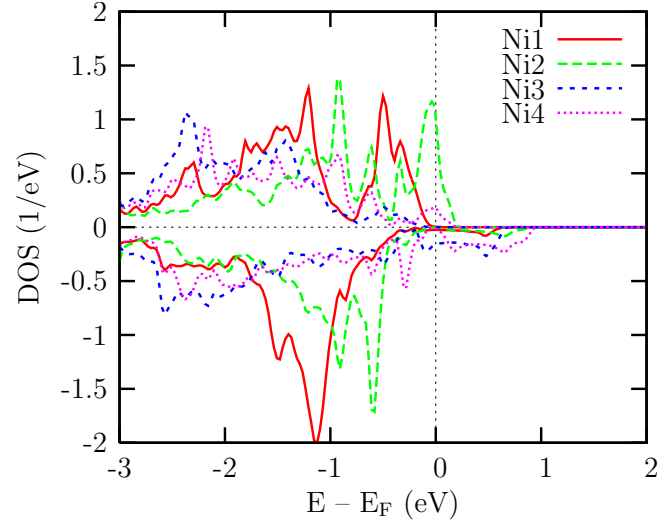}
\caption
{
(Color online) 
Partial spin-majority and spin-minority Ni $3d$ DOS (per Ni atom)
for spin-polarized La$_{5/3}$Sr$_{1/3}$NiO$_4$ in the
CDFAS phase, as obtained by the LDA+$U$ method.
}
\label{fig:fig3}
\end{figure}

Introduction of the spin-polarization has serious effects on the DOS of the magnetic
Ni atoms. In figs.\ \ref{fig:fig2} and \ref{fig:fig3} we address the spin-polarized
partial Ni $3d$ DOS, separated into contributions from the four crystallographically inequivalent Ni
sites, for two possible antiferromagnetic spin patterns of La$_{5/3}$Sr$_{1/3}$NiO$_4$.
The magnetic structure of the ADFAS phase (fig.\ \ref{fig:fig2}) is
based on a ferromagnetic arrangement of domain walls and magnetic domains, an intraplane
ferromagnet, and antiferromagnetic coupling between adjacent planes. On the contrary, in
the case of the CDFAS (fig.\ \ref{fig:fig3}) the diagonal stripes are ordered
antiferromagnetically within the NiO planes, as depicted in fig.\ \ref{fig:fig0}, whereas the
coupling is ferromagnetic along the perpendicular $c$-axis. For both antiferromagnetic spin
patterns, the DOS shows a broad structure of occupied Ni $3d$ states, and a finite number of
states at the Fermi energy. Lacking of an insulating band gap, however, contradicts
an experimental gap of 0.26\,eV \cite{Che94, Kat96}. In  figs.\ \ref{fig:fig2}
and \ref{fig:fig3} we find broad spin-minority bands extending far beyond the Fermi
level. For the ADFAS phase the same applies to the spin-majority states,
while for antiferromagnetic intraplane coupling these states give rise to a much more
pronounced structure with a remarkable weight at the Fermi level. In addition, we
have tried several other spin patterns in order to obtain an insulating state, but without
any success. In each case, the system appeared to be far from opening a band gap.
Furthermore, we have investigated whether a local interaction on the oxygen sites has to be
taken into account. While the total energy in fact slightly decreases for reasonable
values of $U$ and $J_H$, the DOS shape changes only marginally near the Fermi level.
In a recent LDA+$U$ study on La$_{5/3}$Sr$_{1/3}$NiO$_4$, Yamamoto {\it et al.}
\cite{yamamotoCM} have found an insulating solution with an energy gap of 0.11\,eV.
However, in contrast to our calculations, their data rely on a spin pattern with antiferromagnetic
ordering along the diagonal stripes.

\begin{table*}
\begin{center}
\begin{tabular}{l||c|c|c|c|c}
&DFFS&ADFAS&CDFAS&ADFAS, LDA+$U$&CDFAS, LDA+$U$\\\hline
intraplane coupling&fe&fe&af&fe&af\\
interplane coupling&fe&af&fe&af&fe\\\hline
magn.\ moment Ni1 ($\mu_{\rm B}$)&$0.72$&$0.66$&$1.14$&$1.66$&$1.73$\\
magn.\ moment Ni2 ($\mu_{\rm B}$)&$0.71$&$-0.71$&$1.10$&$-1.53$&$1.63$\\
magn.\ moment Ni3 ($\mu_{\rm B}$)&$0.77$&$0.72$&$-1.02$&1.68&$-1.60$\\
magn.\ moment Ni4 ($\mu_{\rm B}$)&$0.78$&$-0.78$&$-0.89$&$-1.64$&$-1.55$\\\hline
total energy gain (mRyd)&$5.8$&$5.8$&$87$&$152$&$143$
\end{tabular}
\end{center}
\caption{\rm Ni magnetic moments and total energy gain (per Ni atom)
with respect to non-magnetic calculations. Columns 1 to 3 refer to pure
GGA calculations, whereas columns 4 to 6 give the corresponding LDA+$U$
values. The LDA+$U$ results are based on the parameters $U=8$\,eV and $J$=0.8\,eV.}
\label{tab1}
\end{table*}

We next address the stability of the spin patterns under consideration.
According to table \ref{tab1}, all magnetic solutions are
lower in energy than the paramagnetic phase, already for $U=0$.
In particular, the DFFS and ADFAS phases turn out to be almost
degenerate, both revealing an energy gain of 5.8\,mRyd per Ni atom as compared
to the non-magnetic solution. Because these patterns only differ by their spin
arrangement along the $c$-axis, we conclude that the magnetic interlayer
coupling is very weak. Turning to the comparison of the ADFAS and CDFAS
configurations, we obtain that the latter has a substantially lower energy
than the former (6.7\,mRyd per Ni atom). This energy gain can safely be attributed to the
intraplane magnetic coupling, due to the nearly vanishing interplane interaction,
as mentioned above. Remarkably, the evaluation of the local magnetic moments
self-consistently influences the various hopping matrix elements and crystal
fields involved in the calculation. It therefore is unlikely to design a
realistic model without accounting for this self-consistency. Nevertheless,
a large number of calculations for the two-band Hubbard model clearly
point at competing phases, see \cite{pssb,jpcm} and the references therein.

Regarding the magnetic moments, we find comparable amplitudes for all Ni
sites, independent of the magnetic structure. Remarkably, even in the
case of the DFFS configuration the magnetic moments show a strong tendency
to stripe formation. Their values increase substantially
in the CDFAS phase, see table \ref{tab1}, resulting in a lower
energy. Note that the total magnetic moment is finite in one unit cell,
amounting to 0.35\,$\mu_B$ per Ni site. Nevertheless, the moment is expected to vanish
if the supercell would be doubled in the $c$-direction. Since the interlayer
coupling is almost negligible, the result of such a calculation would be
closely related to the present data, yet producing a vanishing total
magnetization. Our calculation thus captures all
relevant features of the CDFAS phase. Moreover, a solution with vanishing
intraplane magnetic moment could not be obtained, and therefore is likely to
be either higher in energy or not to exist.

Concerning the magnetic state
of the Ni ions, we find a superposition of $S_z=1/2$
(Ni$^{3+}$-like) with $S_z=0$  and $S_z=1$ states (Ni$^{2+}$-like, one electron
in each $e_g$ orbital with parallel/antiparallel spin), whereas states involving
doubly occupied orbitals are suppressed \cite{Lamboley}. 
When the onsite electron-electron interaction is included ($U=8$\,eV, $J_H=0.8$\,eV),
the local magnetic moments increase significantly, to about 1.6$\mu_B$ in
both antiferromagnetic phases. This modification arises from a combined effect
of $U$ and $J_H$, as suggested by model calculations \cite{Lamboley}. However,
the increase is amplified in the ADFAS phase as compared to the CDFAS phase,
favoring the former by 9\,mRyd per Ni atom. Thus,
the energy order of the antiferromagnetic patterns is inverted due to the onsite
interaction, hence no longer reflecting the experimental situation. This
fact indicates strong interplay between the magnetic structure, the onsite
Coulomb interaction, and the energetical sequence of the magnetic phases. We
have verified that the latter is not sensitive to (i) the exact
values of $U$ and $J_H$, in a reasonable range compared to experimental
data, and (ii) the inclusion of a local interaction on the O sites.
Combining the above with the absence of an insulating ground state in our band
structure calculation leaves little room for an explanation of the deviation
from the experimental observation: one appealing scenario is provided by the
fact that the final determination of the crystal structure, in particular
the positional parameters of all atoms, has not been performed so far. The
structural estimates used for the atomic positions can be inadequate, suggesting to
perform a structure optimization. Up to now, however, a realistic structure
optimization could not be realized due to a large number of lattice degrees
of freedom. As structural relaxation is expected to be very sizeable for NiO$_6$
octahedra, modification of the Ni-O bonding could easily alter the electronic
structure in the vicinity of the Fermi energy, such that an insulating gap
arises. Of course, this might correct the energy sequence of the spin patterns.

In summary, we have presented a series of spin-resolved electronic structure calculations
for the striped nickelate La$_{5/3}$Sr$_{1/3}$NiO$_4$, including onsite Coulomb interaction.
In contrast to experimental findings, the LDA+$U$ approach results
in A-type diagonal filled antiferromagnetic stripes, while in the pure GGA scheme
C-type diagonal filled antiferromagnetic stripes are favored. Our results indicate
that accordance with the experimental
situation calls for further determination of the structural parameters, since
NiO$_6$ octahedra are known to be subject to strong structural distortions
in many cases. We expect that the same applies to all commensurate doping levels.

\acknowledgments
We thank V.\ Eyert, T.\ Kopp, and  M.\ Raczkowski for helpful
discussions, and acknowledge financial support by the Deutsche Forschungsgemeinschaft (SFB 484) and the
Bayerisch-Franz\"osische Hochschulzentrum.

\end{document}